\documentstyle[eqsecnum,aps,prd,eclepsf]{revtex}
\def\and{{$\&$ }}

\def\gsim{\mathrel{\mathpalette\oversim>}}
\def\oversim#1#2{\lower0.2ex\vbox{\baselineskip0pt\lineskip0pt
  \lineskiplimit0pt\ialign{$#1\hfil##\hfil$\crcr#2\crcr\sim\crcr}}}
\begin{document}
\draft


\parbox{\hsize}{
\begin{flushright}
HUPD-9708,   KUNS-1447 
\end{flushright}

\title{Cosmological Baryon Sound Waves 
Coupled with the Primeval Radiation}
%
%
\author{Kazuhiro Yamamoto}
\address{Department of Physics, Hiroshima University,
  Higashi-Hiroshima 739, Japan}
\author{Naoshi Sugiyama and Humitaka Sato}
\address{Department of Physics, 
Kyoto University,
Kyoto 606-01, Japan}
\date{\today, Phys.Rev.D in press}
\maketitle
}


\vspace{10pt}

\begin{abstract}
  The fluid equations for the baryon-electron system in an expanding
  universe are derived from the Boltzmann equation.  The effect of the
  Compton interaction is taken into account properly in order to
  evaluate the photon-electron collisional term.  As an
  application, the acoustic motions of the baryon-electron system after
  recombination are investigated.  The effective adiabatic index
  $\gamma$ is computed for sound waves of various wavelengths, 
  assuming the perturbation amplitude is small.  The oscillations
  are found to be dumped when $\gamma$ changes from 
  between 1 (for an isothermal process) to 5/3 (for an adiabatic process).
\end{abstract}


\section{Introduction}

One of the biggest interest in astrophysics is to understand 
how structures in the universe are formed.  
Observations reveal that the universe is not simply
homogeneous and isotropic but contains rich structures
from stars to the large scale structure such as clusters 
of galaxies, super-clusters, the great wall and voids in it.  
Although the gravity itself does not have any 
particular scale, several physical scales 
associated with the structure formation, such as the 
Jeans scale, the Silk damping scale and 
the horizon scale at the matter-radiation equality epoch,
are derived by the inclusion of an effect of 
primeval radiation \cite{silk,py,smt}. 
Among them, the Jeans scale determines whether small density
fluctuations can grow against pressure or not.  

The linear evolution of small density perturbations has been well
understood.  Outside the Jeans scale, their evolution is described by
the growing mode solution \cite{wein,Peeb1}.  
Assuming the spherical symmetry, we can follow the
evolution of over-dense regions.  The critical threshold of
a density contrast to collapse 
is $1.69$ in the Einstein-de Sitter universe 
and we can estimate the fraction of the collapsed mass
by employing the extrapolation of the  linear perturbation theory 
\cite{sphr}.  
If the wave length of fluctuations is smaller than the Jeans scale,
however,  fluctuations cannot grow up to the critical threshold 
but begin to oscillate as an acoustic wave.  
Therefore an investigation of the time evolution of the Jeans scale is 
crucial in particular for the study what is the first collapsed 
object in the universe.  

The Jeans scale is ruled by the sound speed $c_{\rm s}$ and 
the free fall time of the object or the expansion time of 
the universe. 
After the recombination of hydrogen atoms, the radiation pressure 
becomes ineffective and equations of state are described by the
adiabatic index $\gamma$ of baryonic matter.  If the energy 
transfer between the baryon and the photon is efficient, 
we expect isothermal process, i.e., $\gamma =1$.  
However, if the time scale of the energy transfer 
becomes longer than the oscillation time scale of the sound wave, 
the adiabatic
$\gamma =5/3$ must be achieved.  These two $\gamma$'s give about 
factor two difference for the  Jeans scales in mass.  
This point is crucial when we investigate the evolution of the
baryon density perturbations smaller than the Jeans scale after 
recombination.
Moreover, the temperature of gas is very close to the 
radiation temperature at the recombination epoch but 
gradually separates from it.  Eventually, it does change 
the dependence on the red-shift $z$ from $1+z$ to $(1+z)^2$.  
In order to obtain an accurate Jeans scale,   
we need to take into account all these details.
In this paper, we formulate the energy transfer between 
baryons, electrons and photons from the first principle, 
i.e., the Boltzmann equation. 
Similar treatment on photons has been done by Hu, Scott \& Silk
\cite{HSS} and Dodelson \& Jubas \cite{DJ}.  These works focused on 
the perturbations of photons which are massless particles 
in order to 
get anisotropies of the cosmic microwave background radiation.   
Here we formulate perturbation equations of baryon-electron fluid
which is the massive particle system.  The energy transfer between 
this fluid system and photons is coming through the Compton scattering.
We study in detail the sound speed and the Jeans scale 
after the recombination.
It is found that damping of the sound wave is efficient during the
change of $\gamma$.

In \S II, we describe the Boltzmann equation 
in the perturbed expanding universe. In \S III,
the equations of the baryon-electron system
are derived based on the fluid approximation,
by integrating the Boltzmann equation 
for the momentum space.
The integrations of the collisional term
which describes the Compton interaction
between electrons and photons is 
summarized in \S IV.
The perturbation equations of the baryon-electron
fluid are obtained in \S V.
In \S VI, we investigate the acoustic motion of 
the baryon-electron fluid after the recombination
in the expanding universe as an application of 
our formalism. \S VII is devoted to summary and discussions.
In appendix A, we consider the perturbation equation
for the rate equation in order to complete the 
perturbed equations obtained in \S V.
In appendix B, we put summary of the physical scales
for the cosmological baryon perturbations in the universe.
The evolution of the matter temperature
is summarized in appendix C.

We will work in units where $c=\hbar=k_B=1$.

\section{Boltzmann Equation}
\def\aova{{a\over a_0}}
\def\falpha{{f_{(\alpha)}}}
\def\malpha{{m_{(\alpha)}}}
\def\nalpha{S{n_{(\alpha)}}}
\def\vecx{{x^i}}
\def\vecq{{q^i}}
\def\vecp{{p^i}}
\def\bfx{{\bf x}}
\def\aeq{{a_{\rm eq}}}
\def\Msolar{{M_{\odot}}}
\def\N{N}
\def\aeq{{a_{\rm eq}}}
\def\keq{{k_{\rm eq}}}
\def\Th{\Theta_{2.726}}
\def\xe{{x_e}}
\def\nele{{n_{\rm e}}}
\def\fnu{{f_\nu}}
\def\The{\Theta_{2.726K}}
\def\yp{{y_{\rm p}}}
\def\nb{{n_{\rm b}}}
\def\MJ{{M_{\rm J}}}
\def\Mpc{{\rm Mpc}}
\def\calA{{\cal A}}
\def\kphys{{k^{\rm phys}}}
\def\kcomv{{k^{\rm comv}}}
\def\rhob{{\rho_{\rm b}}}
\def\Omegab{{\Omega_{\rm b}}}
\def\Omegam{{\Omega_{\rm 0}}}
\def\Te{{T_{\rm e}}}
\def\cf{{c_{\rm f}}}
\def\cs{{c_{\rm s}}}
\def\ce{{c_{\rm e}}}
\def\me{{m_{\rm e}}}
\def\aova{{a\over a_0}}
We write the perturbed space-time to the Friedmann-Robertson-Walker
space-time in the Newtonian gauge as 
\begin{equation}
  ds^2=g_{\mu\nu}dx^\mu dx^\nu = -(1+2\Psi)dt^2
  +\biggl(\aova\biggr)^2 (1+2\Phi)\delta_{ij}dx^idx^j ~ ,
\label{metric}
\end{equation}
with introducing the perturbed gravitational potential $\Psi$ and 
the curvature
perturbation $\Phi$.  $a$ is the scale factor and suffix $0$ indicates
the present value.  
Note that we employ a flat orthogonal coordinate system 
besides the scale factor as indicated by the Kronecker's delta
$\delta_{ij}$ for a background space.  
$\Psi$ satisfies the Poisson equation
\begin{equation}
\nabla^2\Psi = 4\pi G \rho {\left(a \over a_0\right)}^2  \delta ~ ,
\end{equation}
where $\rho$ and $\delta$  are total background density and 
density fluctuation, respectively,  and
$\Phi = -\Psi $ 
when the anisotropic stress is negligible, e.g., in the
matter-dominated era.  

We write next the Boltzmann equation for the
distribution function $\falpha(t,\vecx,\vecq)$ of the
$(\alpha)$-particle as 
\begin{equation}
  {\partial\falpha\over\partial t}
  +{\partial\falpha\over\partial\vecx}{d\vecx\over dt}
  +{\partial\falpha\over\partial\vecq}{d\vecq\over dt}
  =C[\falpha] ~ ,
\label{BoltzmannC}
\end{equation}
with the collision term in the right hand side.
This Boltzmann equation is written in terms of the momentum 
$\vecq$ measured by an observer in the cosmological rest frame.
In order to solve this equation, we must rewrite the terms
$d \vecx/ d t$ and $d \vecq/ d t$ in terms of $\vecx$ and $\vecq$, 
which is obtained from the equation of motion of the $(\alpha)$-particle.
To obtain this equation, we write the 
4-momentum in the locally orthonormal frame as $(q^0,\vecq)$.
Here the energy in this frame $q^0$ is defined as
\begin{equation}
  q^0:=\sqrt{{\bf q}^2+\malpha^2},
\end{equation}
where ${\bf q}^2=\sum_i (q^i)^2$. 

The equation of motion is given from the geodesic equation.
However, the geodesic equation of the $(\alpha)$-particle is commonly
written in terms of the 4-momentum $p^\mu$ in the frame (\ref{metric}),
which is defined by $p^\mu=dx^\mu/d\lambda$. 
Here $\lambda(d\lambda=ds/m_{(\alpha)})$ is the affine parameter.
We use the fact that the 4-momentum $(q^0,\vecq)$ is related to $p^\mu$, 
as follows,
\begin{eqnarray}
  &&\vecq=\sqrt{g_{ii}}{d x^i\over d\lambda}=\aova (1+\Phi)\vecp ~ ,
\label{relvecq}
\\
  &&q^0=\sqrt{-g_{00}}{d x^0\over d\lambda}=(1+\Psi)p^0 ~ .
\label{relqzero}
\end{eqnarray}
Equations (\ref{relvecq}) and (\ref{relqzero}) give the following
relations, up to the first order of $\Psi$ and $\Phi$, 
\begin{equation}
  {d\vecx\over dt}={\vecp\over p^0}
  ={a_0\over a}(1+\Psi-\Phi){\vecq\over q^0} ~ ,
\label{dxdt}
\end{equation}
and
\begin{equation}
  {dq^i \over dt}={\dot a \over a} \vecq
  +\aova\biggl({\partial \Phi\over\partial t}
    +{\partial \Phi\over\partial x^j}{d x^j\over dt}\biggr)\vecp
  +\aova(1+\Phi){d \vecp\over dt},
\label{dqovdt}
\end{equation}
where the over-dot denotes $t$-differentiation.

On the other hand, the geodesic equation in the leading order of the
perturbation derives
\begin{equation}
  {d\vecp\over dt}=-\biggl({a_0\over a}\biggr)^2
  \biggl[2{a\over a_0}{\dot a\over a}(1-\Phi)\vecq
        +2{a\over a_0}\dot\Phi\vecq
        +2\Phi_{,j}{q^jq^i\over q^0}+
        \delta^{ij}\left( \Psi_{,j} q^0 
        - \Phi_{,j} {{\bf q}^2\over q^0}\right) \biggr],
\end{equation}
where $_{,i}=\partial /\partial x^i$.  Hereafter we will omit $\delta^{ij}$
and will write as e.g., $\Psi_{,i} = \delta^{ij}\Psi_{,j}$.
Inserting this to Eq.(\ref{dqovdt}),
\begin{equation}
  {dq^i \over dt}=-{a_0\over a}\biggl[
  {a\over a_0}{\dot a\over a}\vecq+{a\over a_0}\dot\Phi\vecq
  +\Phi_{,j}{q^jq^i\over q^0}+\Psi_{,i}q^0-\Phi_{,i}{{\bf q}^2\over q^0}
  \biggr].
\label{dqovdtb}
\end{equation}
Now we can write down the left hand side of the Boltzmann equation
(\ref{BoltzmannC})
using the equations (\ref{dxdt}) and (\ref{dqovdtb}). If we employ the
conformal time defined by $(a/a_0)d\eta \equiv dt$
instead of the proper time,
it becomes as
\begin{eqnarray}
  &&{\partial \falpha\over \partial\eta}
  +{\partial \falpha\over \partial \vecx}(1-\Phi+\Psi){\vecq\over q^0}
\nonumber
\\
&&\hspace{1.0cm}
  +{\partial \falpha\over \partial \vecq}
  \biggl[-{a'\over a}\vecq-\Phi'\vecq-\Psi_{,i}q^0
         -\Phi,j{q^jq^i\over q^0}+\Phi_{,i}{{\bf q}^2\over q^0}\biggr]
  =\aova C[\falpha],
\label{Boltz}
\end{eqnarray}
where the prime denotes $\eta$-differentiation.
Until now,  we have assumed only the
smallness of  perturbations to the homogeneous background.

If we introduce a further assumption that the motion of the 
$(\alpha)$-particle is non-relativistic,  
the Boltzmann equation (\ref{Boltz}) reduces to
\begin{equation}
  {\partial \falpha\over \partial\eta}
  +{\partial \falpha\over \partial \vecx}{\vecq\over \malpha}
  +{\partial \falpha\over \partial \vecq}
  \biggl[-\biggl({a'\over a}+\Phi'\biggr)\vecq-\Psi_{,i}\malpha\biggr]
  =\aova C[\falpha] ~ ,
\label{BoltzmannB}
\end{equation}
where $v^i=\vecq/\malpha $ and terms of $O({\bf v}^2 \times 
\Psi ~{\rm or} ~\Phi)$ and $O({\bf v}^3)$ are omitted. 
This is a familiar non-relativistic Boltzmann 
equation in the expanding universe, when the term proportional 
to $\Phi'$ is neglected.

\section{Fluid Approximation}
\def\vb{{v_{\rm b}}}
\def\bfvb{{{\bf v}_{\rm b}}}
\def\Te{{T_{\rm b}}}
\def\bfqalpha{{{\bf q}_{(\alpha)}}}
\def\rhob{{\rho_{\rm b}}}
\def\me{{m_{\rm e}}}
We consider a system of non-relativistic particles (baryons and electrons)
and photons which interact only with the electrons.
Neutral and ionized hydrogen atoms and neutral helium atoms are 
taken into account as baryonic components. 
We further take 
a single fluid approximation for this baryon and electron system since 
the time scale of the interaction between them is short enough.
Under these assumptions, we here derive the fluid equation from the 
Boltzmann equation (\ref{BoltzmannB}).

Since the particles are non-relativistic, we take the distribution
function of 
the $(\alpha)$-particle, where $\alpha =\rm{ e, H}$ and $\rm He$ for 
the electron, the hydrogen and the helium,
as 
\begin{equation}
  \falpha=\nalpha(\bfx)\biggl({2\pi\over\malpha \Te(\bfx)}\biggr)^{3/2}
  \exp\biggl[{-(\bfqalpha-\malpha\bfvb(\bfx))^2\over2\malpha
  \Te(\bfx)}\biggr]  ,
\label{distrib}
\label{formoff}
\end{equation}
with a normalization of 
\begin{equation}
  \int{d^3\bfqalpha\over(2\pi)^3} \falpha=\nalpha(\bfx) ~ .
\end{equation} 
Here $\Te(\bfx)$ and $\bfvb(\bfx)$ are 
the temperature and the peculiar velocity of this fluid system  
which the suffix $b$ denotes.
Then it follows that 
\begin{eqnarray}
  &&\int{d^3\bfqalpha\over(2\pi)^3} q^i_{(\alpha)} \falpha =
  \malpha \nalpha \vb ^i,
\\
  &&\int{d^3\bfqalpha\over(2\pi)^3} q^i_{(\alpha)} q^j_{(\alpha)} \falpha =
  \malpha^2 \nalpha \vb^i \vb^j +\malpha\nalpha \Te \delta^{ij},
\\
  &&\int{d^3\bfqalpha\over(2\pi)^3} \bfqalpha^2 q^i_{(\alpha)} \falpha =
  \malpha^3  \nalpha \bfvb^2 \vb^i  + 5 \malpha^2 \nalpha \Te \vb^i ~ .
\end{eqnarray}

Operating the following integration and the summation for the Boltzmann 
equation (\ref{BoltzmannB})
with the distribution function (\ref{formoff}),
\begin{equation}
  \sum_{(\alpha)} \malpha \int{d^3\bfqalpha\over(2\pi)^3} ~ ,
\end{equation}
we obtain the continuity equation
\begin{equation}
  {\partial\rhob\over\partial\eta}+3\biggl({a'\over a}+\Phi'\biggr)\rhob
  +{\partial\over \partial x^i}\Bigl( \rhob \vb^i \Bigr)=0
\label{contA}
\end{equation}
for
\begin{equation}
  \rhob(\bfx) \equiv \sum_{(\alpha)} \malpha\nalpha(\bfx).
\end{equation}
The collision terms should cancel out by the summation.

Operating the following integration and the summation to Eq.(\ref{BoltzmannC}),
\begin{equation}
  \sum_{(\alpha)} \int{d^3\bfqalpha\over(2\pi)^3} ~ q_{(\alpha)}^i,
\end{equation}
we get the Euler equation
\begin{equation}
  {\partial(\rhob\vb^i)\over\partial\eta}+4\biggl({a'\over a}+\Phi'\biggr)\rhob\vb^i
  +{\partial(\rhob\vb^i\vb^j)\over\partial x^j}
  +{\partial P\over\partial x^i}
  +\rhob{\partial \Psi\over\partial x^i}
  =\Delta V^i_{\rm Compton} ~ ,
\label{momnA}
\end{equation}
where 
\begin{equation}
  P\equiv\sum_{(\alpha)}\nalpha \Te,
\end{equation}
and
\begin{equation}
  \Delta V^i_{\rm Compton}\equiv\aova
  \int{d^3{\bf q}_{\rm (e)} \over(2\pi)^3}~q_{({\rm e})}^i
  C[f_{(\rm e)}]_{\rm Compton} ~ .
\end{equation}
The contributions from the collision term between the baryon and 
the electron should cancel out while the contribution from the Compton 
interaction between the electron and the photon, $\Delta V_{\rm Compton}$,
remains. The explicit form of $\Delta V^i_{\rm Compton}$
is given in the next section.

Operating the following integration and the summation to Eq.(\ref{BoltzmannC}),
\begin{equation}
  \sum_{(\alpha)} {1\over 2\malpha} 
  \int{d^3\bfqalpha\over(2\pi)^3} ~ {\bf q}_{(\alpha)}^2 ~ ,
\end{equation}
we get the energy equation
\begin{eqnarray}
  &&{\partial\over\partial\eta}
  \biggl(\rhob\biggl({\bfvb^2\over2}+h\biggr)\biggr)
  +5\Bigl({a'\over a}+\Phi'\Bigr)\rhob\biggl({\bfvb^2\over2}+h\biggr)
  +{\partial\over\partial x^i}
  \biggl(\vb^i\rhob\biggl({\bfvb^2\over2}+h\biggr)\biggr)
\nonumber
\\
  && 
  -{\partial P\over\partial\eta}-5\Bigl({a'\over a}+\Phi'\Bigr)P
  +{\partial \Psi\over\partial x^i}\vb^i \rhob=\Delta E_{\rm Compton} ~ ,
\label{enegA}
\end{eqnarray}
where
\begin{equation}
  h \equiv {5 P\over 2\rhob}
  ={5 \sum_{(\alpha)}\nalpha \Te \over 2\sum_{(\alpha)} \malpha\nalpha}
  ~ ,
\end{equation}
and
\begin{equation}
  \Delta E_{\rm Compton}\equiv {1\over 2\me} \aova
  \int{d^3 {\bf q}_{\rm (e)}\over(2\pi)^3} ~ {\bf q}_{\rm (e)}^2
  C[f_{(\rm e)}]_{\rm Compton} ~ . 
\end{equation}

\section{Collision Term}
\def\fgamma{{f_{(\gamma)}}}
\def\felect{{f_{(\rm e)}}}
\def\ne{{n_{\rm e}}}
\def\sigmaT{{\sigma_{\rm T}}}
\def\bfp{{\bf p}}
\def\bfq{{\bf q}}
\def\Tgamma{{T_\gamma}}
Here let us evaluate the collision term between 
the electron and the photon. 
The explicit form of $C[f_{(\rm e)}]_{\rm Compton}$ is
\begin{eqnarray}
  &&C[f_{(\rm e)}(\bfq)]_{\rm Compton}=
\nonumber
\\
  &&\hspace{1cm}
  {1\over 2 q^0} \int\int\int{2 d^3\bfp\over (2\pi)^3 2 p^0}
  {2 d^3\bfp'\over (2\pi)^3 2 p'^0}
  { 2d^3 {\bf q}'\over (2\pi)^3 2 q'^0}
  (2\pi)^4\delta^{(4)}(p+q-p'-q') \bigl\vert M\bigr\vert^2
\nonumber
\\
  &&\hspace{1cm}
  \times\biggl\{
  (1+\fgamma(\bfp))\fgamma(\bfp')\felect(\bfq')
  -(1+\fgamma(\bfp'))\fgamma(\bfp)\felect(\bfq)  \biggr\},
\label{collisionA}
\end{eqnarray}
where $\fgamma(\bfp)$ is the photon distribution function, 
$\bf p$ and $\bf p'$ are the photon momenta,
$p^0\equiv\big\vert \bf p\big\vert$, 
$p'^0\equiv\big\vert \bf p'\big\vert$ and 
$\delta^{(4)}(p+q-p'-q')=\delta(p^0+q^0-p'^0-q'^0)
\delta^{(3)}(\bf p+q-p'-q')$ 
is the 4-dimensional Dirac's delta function.
$\bigl\vert M\bigr\vert^2$ is the matrix element 
summing and averaging over the electron spin and the photon
polarization \cite{MS} which is described as,
\begin{equation}
  \bigl\vert M\bigr\vert^2={(4\pi)^2\over 2} \alpha_{\rm EM}^2 
  \biggl({\tilde p'\over \tilde p}
  +{\tilde p\over \tilde p'}-\sin^2\tilde\beta\biggr).
\end{equation} 
Here $\tilde p$ and $\tilde \beta$ are the photon energy and
the photon scattering angle, respectively, in the electron 
rest frame, and $\alpha_{\rm EM}^2=3\me^2\sigmaT/8\pi$.

Since the explicit integration to get the collision term is 
very tedious,
instead of performing the integration of (\ref{collisionA}), 
we here consider the Boltzmann equation for the photon distribution
function
\begin{equation}
  {\partial\fgamma\over\partial t}
  +{\partial\fgamma\over\partial\vecx}{d\vecx\over dt}
  +{\partial\fgamma\over\partial p^i}{d p^i\over dt}
  =C[\fgamma]_{\rm Compton} ~ ,
\end{equation}
where
\begin{eqnarray}
  &&C[\fgamma(\bfp)]_{\rm Compton}=
\nonumber
\\
  &&\hspace{1cm}
  {1\over 2 p^0} \int\int\int{2 d^3\bfq\over (2\pi)^3 2 q^0}
  {2 d^3\bfq'\over (2\pi)^3 2 q'^0}{ 2d^3 \bfp'\over (2\pi)^3 2 p'^0}
  (2\pi)^4\delta^{(4)}(p+q-p'-q') \bigl\vert M\bigr\vert^2
\nonumber
\\
  &&\hspace{1cm}
  \times{1\over 2}\biggl\{
  (1+\fgamma(\bfp))\fgamma(\bfp')\felect(\bfq')
  -(1+\fgamma(\bfp'))\fgamma(\bfp)\felect(\bfq)  \biggr\}.
\label{collisionB}
\end{eqnarray}
Note the factor of $1/2$ in the last line. We should put 
this factor because we have defined the electron distribution 
function by Eq.(\ref{distrib}) or Eq.(\ref{distribe}), 
where the spin degree of freedom is summed.

The integration of the collision term Eq.(\ref{collisionB}) 
has been performed by Dodelson \& Jubas \cite{DJ} and 
Hu, Scott \& Silk \cite{HSS}.
The calculation is based on the assumption that the electron motion is
non-relativistic, and is performed by expanding in terms of $O(1/\me)$.
In particular, in the paper by Dodelson \& Jubas \cite{DJ}, 
the integration of the collision term is carried
by assuming the same form of the electron distribution function
as Eq.(\ref{distrib}), i.e.,
\begin{equation}
  \felect=\ne(\bfx)\biggl({2\pi\over\me \Te(\bfx)}\biggr)^{3/2}
  \exp\biggl[{-(\bfq_{\rm e}-\me\bfvb(\bfx))^2\over2\me
  \Te(\bfx)}\biggr] ,
\label{distribe}
\end{equation}
and by expressing the photon distribution function
in the power expansion of $O(1/\me)$ up to the second order
\begin{equation}
   \fgamma=\fgamma^{(0)}(p^0)+\fgamma^{(1)}(\bfp)+\fgamma^{(2)}(\bfp)  ~ .
\end{equation}
Following their result \cite{DJ}
\footnote{Eq.(5.1) in their paper seems to contain a typographical error.},
the collision term is written in the second order of perturbation as 
\begin{equation}
C[\fgamma(\bfp)]_{\rm Compton}=C^{(1)}+C^{(2)},
\end{equation}
with
\begin{equation}
  C^{(1)}=\ne\sigmaT\biggl[\fgamma_0^{(1)}+{1\over2}\fgamma_2^{(1)}
  {\rm P}_{2}(\mu)-\fgamma^{(1)}-p^0{\partial\fgamma^{(0)}\over\partial p^0}
  \mu\vb\biggr],
\end{equation}
and 
\begin{eqnarray}
  &&C^{(2)}=\ne\sigmaT\biggl[\fgamma_0^{(2)}+{1\over2}\fgamma_2^{(2)}{\rm P}
  _2(\mu)-\fgamma^{(2)}
\nonumber
\\
&&\hspace{2.0cm}
  +\vb^2p^0{\partial\fgamma^{(0)}\over\partial p^0}
  (\mu^2+1)+\vb^2(p^0)^2
   {\partial^2\fgamma^{(0)}\over(\partial p^0)^2}
   ({11\over20}\mu^2+{3\over20})
\nonumber
\\
&&\hspace{2.0cm}+{1\over\me (p^0)^2} {\partial\over\partial p^0} 
\biggl\{ (p^0)^4 \biggl(
   \Te{\partial\fgamma^{(0)}\over\partial p^0}+\fgamma^{(0)}(1+
   \fgamma^{(0)})\biggr) \biggr\} \biggr],
\end{eqnarray}
where 
\begin{eqnarray}
  &&\fgamma^{(i)}_l=\int_{-1}^{1}{d\mu\over2}
  {\rm P}_{l}(\mu) \fgamma^{(i)},
\\
  &&\mu={\bfvb\cdot\bfp\over 
  \big\vert\bfvb\big\vert\big\vert\bfp\big\vert },
\end{eqnarray}
with ${\rm P}_{l}(\mu)$ being a Legendre polynomial.

The energy transfer rate is described as 
\begin{equation}
  \Delta E_{\rm Compton}=
  -{\aova}2\int{d^3\bfp\over(2\pi)^3} 
  p^0 C[\fgamma(\bfp)]_{\rm Compton} ~ .
\end{equation}
Therefore we obtain 
\begin{equation}
  \Delta E_{\rm Compton}=-
  {\aova}{4\ne\sigmaT\rho_{\gamma^{(0)}}\over \me}
  \biggl({\me\over3}\bfvb^2(\bfx)-\Tgamma^{(0)}+\Te(\bfx)\biggr),
\end{equation}
where $\Tgamma^{(0)}$ is the background photon temperature
and $\rho_\gamma^{(0)}=\pi^2\Tgamma^{(0)4}/15$ is the background photon 
energy density.

The momentum transfer rate can be found from 
\begin{equation}
  \Delta V_{\rm Compton}=
  -{\aova}2\int{d^3\bfp\over(2\pi)^3} \bfp C[\fgamma(\bfp)]_{\rm
  Compton} ~ , 
\end{equation}
and the result of the integration in the leading order of 
perturbations gives the well known form \cite{Peeb1},
\begin{equation}
  \Delta V_{\rm Compton}=
  {4\over3}{\aova}{\ne\sigmaT\rho_\gamma^{(0)}}\Bigl({\bf v}_\gamma-\bfvb \Bigr).
\end{equation}

\section{Perturbation Equations}
\def\nb{{n_{\rm B}}}
\def\mp{{m_{\rm p}}}
\def\alphaeq{{x_{\rm eq}}}
\def\yp{{y_{\rm p}}}
\def\deltab{{\delta_{\rm b}}}
Electrons({\rm e}), neutral and ionized hydrogen atoms({\rm H}) and 
helium atoms({\rm He})
are the particle species of the baryon-electron fluid.
The number densities for each species are written
\begin{eqnarray}
  &&n_{\rm e}=x\Bigl(1-{\yp\over2}\Bigr)\nb,
\label{nenb}
\\
  &&n_{\rm H}=\Bigl(1-{\yp}\Bigr)\nb,
\\
  &&n_{\rm He}={\yp\over4}\nb,
\end{eqnarray}
where 
$\yp$ is the primordial helium mass fraction,
$n_{\rm H}$ and $n_{\rm He}$ are the number density of 
neutral and ionized hydrogen and helium, respectively,  
$\nb \equiv n_{\rm H} + 4n_{\rm He}$ is the total baryon number density 
and $x$ is an ionization fraction defined as 
$x \equiv n_{\rm e}/ \left(n_{\rm H} + 2n_{\rm He}\right)$.
Then,
\begin{eqnarray}
  &&\rhob=\sum_{(\alpha)}\malpha\nalpha\simeq m_{\rm p}\nb
\label{mpnb}
\\
  &&P=\sum_{(\alpha)}\nalpha \Te
  =\biggl(1+x-{\yp\over2}\Bigl(x+{3\over2}\Bigr)\biggr)\nb \Te,
\end{eqnarray}
where $\mp$ is the proton mass.

First let us summarize the equations for the baryon-electron
fluid in the expanding universe. 
The continuity equation is (\ref{contA}).
The Euler equation (\ref{momnA}) reduces to
\begin{equation}
  {\partial\vb^i\over\partial\eta}+\biggl({a'\over a}+\Phi'\biggr)\vb^i
  +\vb^j{\partial\vb^i\over\partial x^j}
  +{1\over\rhob}{\partial P\over\partial x^i}
  +{\partial \Psi\over\partial x^i}
  ={\aova}{\ne\sigmaT\over R}\Bigl({v}^i_\gamma-\vb^i \Bigr),
\label{EulerEQ}
\end{equation}
with the use of  Eq.(\ref{contA}) and $R \equiv 3\rhob/4\rho_\gamma^{(0)}$.
The energy equation (\ref{enegA}) reduces to 
\begin{eqnarray}
  &&{\partial\over\partial\eta}\biggl({\bfvb^2\over2}+h\biggr)
  +2\Bigl({a'\over a}+\Phi'\Bigr)\biggl({\bfvb^2\over2}+h\biggr)
  +\vb^i{\partial\over\partial x^i}
  \biggl({\bfvb^2\over2}+h\biggr)
  -{1\over\rhob}{\partial P\over\partial\eta}
  -{5\over\rhob}\Bigl({a'\over a}+\Phi'\Bigr)P
\nonumber
\\
  &&\hspace{0.5cm}
  +{\partial \Psi\over\partial x^i}\vb^i
  =
  4{\aova}{x\Bigl(1-\yp/2\Bigr)\sigmaT\rho_\gamma^{(0)}\over\me\mp}
  \biggl(T_\gamma^{(0)}-\Te(\bfx)-{\me\over3}\bfvb^2\biggr),
\label{energEQ}
\end{eqnarray}
by using Eqs.(\ref{contA}), (\ref{nenb}) and (\ref{mpnb}).
We supplement the equation of state,
\begin{equation}
  h={5 P\over2 \rhob} ~ , 
  \hspace{1cm}
  P={\rhob\over \mp}\biggl(1+x-
  {\yp\over2}\Bigl(x+{3\over2}\Bigr)\biggr)\Te ~ .
\end{equation}

Now we solve the above equations
by a perturbative method assuming a small deviations from the uniformity.
Define the perturbative expansions as follows,
\begin{eqnarray}
  &&\rhob=\rhob^{(0)}(\eta)(1+\deltab(\eta,\bfx)),
\\
  &&h=h^{(0)}(\eta)(1+\Delta_h(\eta,\bfx)),
\\
  &&P=P^{(0)}(\eta)(1+\Delta_P(\eta,\bfx)),
\\
  &&\Te=\Te^{(0)}(\eta)(1+\Delta_\Te(\eta,\bfx)),
\\
  &&x=x^{(0)}(\eta)+\delta x(\eta,\bfx),
\end{eqnarray}
together with $\bfvb=\bfvb(\eta,\bfx)$.

The zero-th order equations of (\ref{contA}) and (\ref{energEQ}) are
\begin{equation}
  {\rhob^{(0)}}'+3{a'\over a}\rhob^{(0)}=0,
\end{equation}
and
\begin{equation}
  {\partial h^{(0)}\over\partial\eta}
  +2{a'\over a}h^{(0)}
  -{1\over\rhob^{(0)}}\biggl({\partial P^{(0)}\over\partial\eta}
  +5{a'\over a}P^{(0)}\biggr)
  =4{\aova}{x^{(0)}\Bigl(1-\yp/2\Bigr)
  \sigmaT\rho_\gamma^{(0)}\over\me\mp}
  \Bigl(T_\gamma^{(0)}-\Te^{(0)}\Bigr),
\label{energyC}
\end{equation}
respectively. Eq.(\ref{energyC}) reduces to \cite{Peeb2}
\begin{equation}
  {\Te^{(0)}}'+2{a'\over a}\Te^{(0)}=\eta_{\rm E}^{-1}
  (T_\gamma^{(0)}-\Te^{(0)}),
\label{tempreq}
\end{equation}
where we have neglected the term proportional to $x^{(0)}{}'\Te^{(0)}$
\footnote{
As we will see in the below, the matter temperature 
follows the photon temperature in the recombination regime.
This is because the energy transfer time scale $\eta_{\rm E}$
is small enough. After recombination, the fraction of the residual 
ionization $x^{(0)}$ is almost fixed, and the time variation of $x^{(0)}$ 
is small. This is the reason why we neglected the term 
proportional to $x^{(0)}{}'\Te^{(0)}$.}  
and used the zero-th order equation of state,
\begin{equation}
  h^{(0)}={5 P^{(0)}\over2 \rhob^{(0)}},
  \hspace{1cm}
  P^{(0)}={\rhob^{(0)}\over\mp}\biggl(1+x^{(0)}-
  {\yp\over2}\Bigl(x^{(0)}+{3\over2}\Bigr)\biggr)\Te^{(0)}.
\end{equation}
We defined the Compton energy transfer time scale $\eta_{\rm E}$ as 
\begin{equation}
  \eta_{\rm E}^{-1}={8\over3}{\aova}{x^{(0)}\Bigl(1-\yp/2\Bigr)
  \sigmaT\rho_\gamma^{(0)}\over\me
  \Bigl(1+x^{(0)}-(x^{(0)}+3/2)\yp/2\Bigr)} ~ .
\end{equation}
The solution of Eq.(\ref{tempreq}) is discussed in appendix C.

The perturbed equations of (\ref{contA}) and (\ref{EulerEQ}) are
\begin{equation}
  \deltab'+3\Phi'+{\partial \vb^i\over \partial x^i}=0,
\label{contEQC}
\end{equation}
and 
\begin{equation}
  {\vb^i}'+{a'\over a}\vb^i+{P^{(0)}\over \rhob^{(0)}} 
  {\partial \Delta_P\over \partial x^i} 
  +{\partial\Psi\over\partial x^i}
  =
  {\aova}{\ne\sigmaT\over R}\Bigl({v}^i_\gamma-\vb^i \Bigr).
\label{EulerEQC}
\end{equation}

The perturbed equation of (\ref{energEQ}) yields
\begin{eqnarray}
  &&\Delta_h'
  +\biggl({5\over3}{{h^{(0)}}'\over h^{(0)}}-{2\over3}{{P^{(0)}}'\over P^{(0)}}\biggr)
  \Delta_h-{2\over3}\deltab'
\nonumber
\\
  &&\hspace{2cm}
  =\eta_{\rm E}^{-1}\biggl(-\Delta_h
  +{\delta x\over x^{(0)}}{T_\gamma^{(0)}-\Te^{(0)}\over\Te^{(0)}}
  +{\Bigl(1-{\yp/2}\Bigr)\delta x\over 
    1+x^{(0)}-(x^{(0)}+3/2){\yp/2}}\biggr),
\label{energEQC}
\end{eqnarray}
where we have used the perturbative part  of the equation of state
as follows, 
\begin{equation}
  \Delta_h=\Delta_P-\deltab ,
\label{eqst1}
\end{equation}
and 
\begin{equation}
  \Delta_P=\deltab+\Delta_{\Te}+{\Bigl(1-{\yp/2}\Bigr)\delta x\over 
    1+x^{(0)}-(x^{(0)}+3/2){\yp/2}} ~ .
\label{eqst2}
\end{equation}
In order to complete the perturbation equations,
we need an equation to specify the evolution 
of the perturbation of the ionization rate $\delta x$
in equations (\ref{energEQC}) and (\ref{eqst2}),
which we have considered in the appendix A.
In the early phase of the recombination 
the terms proportional to $\delta x$ become important.

\section{Sound waves after the recombination}
\def\cf{{c_{\rm f}}}
\def\ce{{c_{\rm e}}}
\def\cs{{c_{\rm s}}}
\def\Ms{{{\rm M}_\odot}}
As an application of the basic relations obtained in the previous
sections, we consider the behavior of 
the sound wave of the baryon-electron fluid 
after the recombination epoch.
Neglecting a perturbation of the gravitational potential, we get
basic equations from Eqs.(\ref{contEQC}), (\ref{EulerEQC}),
(\ref{energEQC}) as 
\begin{eqnarray}
  &&\deltab'+{\partial \vb^i\over\partial x^i}=0,
\label{contAppl}
\\
  &&{\vb^i}'+{a'\over a}\vb^i+{P^{(0)}\over \rhob^{(0)}} 
  {\partial \Delta_P\over \partial x^i} =0,
\label{EulerAppl}
\end{eqnarray}
and 
\begin{equation}
  \Delta_h'
  +\biggl({5\over3}{{h^{(0)}}'\over h^{(0)}}-{2\over3}{{P^{(0)}}'\over P^{(0)}}\biggr)
  \Delta_h-{2\over3}\deltab'=-\eta_{\rm E}^{-1}\Delta_h .
\label{energAppl}
\end{equation}
Here we have also neglected 
a perturbation of the ionization fraction and 
an interaction with the photon 
in the Euler equation
because the momentum transfer rate is so small after the decoupling
The wave equation is derived from Eqs.(\ref{contAppl}) and (\ref{EulerAppl}),  
together with Eq.(\ref{eqst1}), as 
\begin{equation}
  \deltab''+{a'\over a}\deltab'
  -{P^{(0)}\over \rhob^{(0)}}{\partial^2\over\partial\bfx^2}(\Delta_h+\deltab)
  =0.
\label{deltabAppl}
\end{equation}
This must be coupled with the energy equation (\ref{energAppl})
in order to investigate the sound oscillation 
of the baryon-electron fluid.

Now let us omit
the time-dependence of the background quantities.
When the time scale of the sound oscillation is shorter than
the Hubble time, this is a good approximation.
Then Eqs.(\ref{deltabAppl}) and (\ref{energAppl}) reduce
to simple equations
\begin{eqnarray}
  &&\delta_{{\rm b}}(k,\eta)''+{P^{(0)}\over \rhob^{(0)}}
  k^2(\Delta_{h}(k,\eta)+\delta_{{\rm b}}(k,\eta))
  =0,
\label{deltabApplB}
\\
  &&\Delta_{h}(k,\eta)'={2\over3}
  \delta_{{\rm b}}(k,\eta)'-\eta_{\rm E}^{-1}\Delta_{h}(k,\eta),
\label{energApplB}
\end{eqnarray}
where we took the Fourier mode expansion by setting
$\deltab=\delta_{{\rm b}}(k,\eta)e^{i\bf k\cdot x}$,
and $\Delta_h=\Delta_{h}(k,\eta)e^{i\bf k\cdot x}$.
The above equations yield
\begin{equation}
  \eta_{\rm E}{d\over d\eta}
  \biggl({d^2\deltab(k,\eta)\over d\eta^2}
  +\cf^2k^2{\deltab(k,\eta)}\biggr)
  +\biggl({d^2\deltab(k,\eta)\over d\eta^2}
  +\ce^2k^2{\deltab(k,\eta)}\biggr)=0 ~ .
\label{cubiceqA}
\end{equation}
Here we defined 
\begin{equation}
  \cf^2={5\over3}{P^{(0)}\over \rhob^{(0)}},
  \hspace{1cm}
  \ce^2={P^{(0)}\over \rhob^{(0)}}.
\end{equation}
As is well known, 
$\cf$ is the sound speed for the adiabatic state and 
$\ce$ is the one for the isothermal state.

Taking the wave solution 
$\delta_{\rm b}(k,\eta) \propto e^{-i\omega\eta}$, we get
the following dispersion relation,
\begin{equation}
  -i\eta_{\rm E}\omega(\omega^2-\cf^2 k^2)+ \omega^2-\ce^2 k^2=0 ~ .
\label{dispersionA}
\end{equation}
In order to solve this equation, it is convenient to introduce
the variable $\varpi$ such as  $\omega=i\varpi$  
and we have
\begin{equation}
  \ce k \eta_{\rm E} \Bigl({\varpi\over \ce k}\Bigr)
  \biggl( \Bigl({\varpi\over \ce k}\Bigr)^2+{5\over3}\biggr)
  +\Bigl({\varpi\over \ce k}\Bigr)^2+1=0 ~ .
\label{dispersionB}
\end{equation}
The sound speed $\cs$ and the adiabatic index $\gamma$ are
defined as 
\begin{equation}
  \cs=-{\rm Im} \biggl[{\varpi\over k}\biggr],
  \hspace{1cm}
  \gamma={\cs^2\over\ce^2} ~ .
\end{equation}
Therefore we need to solve the cubic equation (\ref{dispersionB})
in order to get the sound speed of the baryon-electron system.

\begin{figure}[t]
  \begin{center}
  \epsfile{file=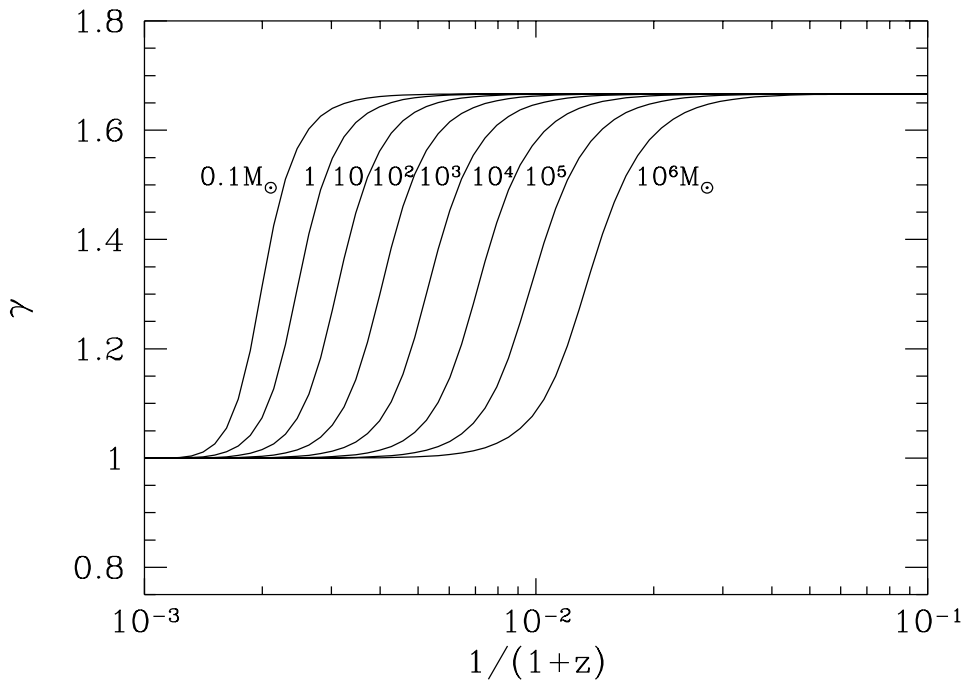,width=13cm}
  \end{center}
\caption{The behaviors of $\gamma$ as a function of $1/(1+z)$. 
Each lines for the mass scales  
$M=1\Ms$, $10^3\Ms$, $10^6\Ms$, $10^9\Ms$, 
$10^{12}\Ms$, respectively, with the cosmological parameters, 
$h=0.5$, $\Omega_0=1.0$, $\Omega_{\rm B}=0.1$.  
$\gamma$ is changed in the earlier stage from $1$ to $5/3$ 
for the smaller scale perturbations.}
\label{Fig.1}
\end{figure}

A cubic equation can be solved exactly in an analytic form
while the expression is complicated. It may be 
instructive to show the solution of Eq.(\ref{dispersionB}) 
in a simple form with some approximation.
Expanding the solution around the adiabatic state, i.e.,
expanding in terms of $\epsilon$ by setting
the solution $\omega/k=\cf(1+\epsilon)$, we find
\begin{equation}
  {\omega\over k}\simeq
  \cf\biggl(1-{1+i k\cf\eta_{\rm E}
  \over5(1+(k\cf\eta_{\rm E})^2)} \biggr).
\label{solapprox}
\end{equation}

It is interesting to clarify
the behavior of the adiabatic index $\gamma$ 
in the expanding universe after the recombination,
which we demonstrate as an example of the usefulness of our 
formalism in the below.
As is clear from Eq.(\ref{dispersionA}),
the adiabatic index $\gamma$ is ruled 
by the ratio of the Compton energy transfer time scale 
$\eta_{\rm E}$ to the sound oscillation time scale.
If $\eta_E \omega \ll 1$, $\omega \simeq c_e k$ and
$\gamma \simeq 1$.  
On the other hand, if 
$\eta_E \omega \gg 1$, $\omega \simeq c_f k$ and
$\gamma \simeq 5/3$. 
Therefore the sound speed depends on scales of the perturbations,
and it is expected that $\gamma$ is changed from $1$ to $5/3$
as the universe expands.

Fig.1 shows $\gamma$ as a function of $1/(1+z)$
by directly solving Eq.(\ref{dispersionB}).
The ionization fraction is calculated properly by 
solving the recombination process in the expanding universe
with the cosmological parameters in the figure caption\cite{JW,HSI}. 
As is expected, $\gamma$ is changed from $1$ to $5/3$ in the earlier 
stage of the universe for smaller scale perturbations. 

\begin{figure}[t]
  \begin{center}
  \epsfile{file=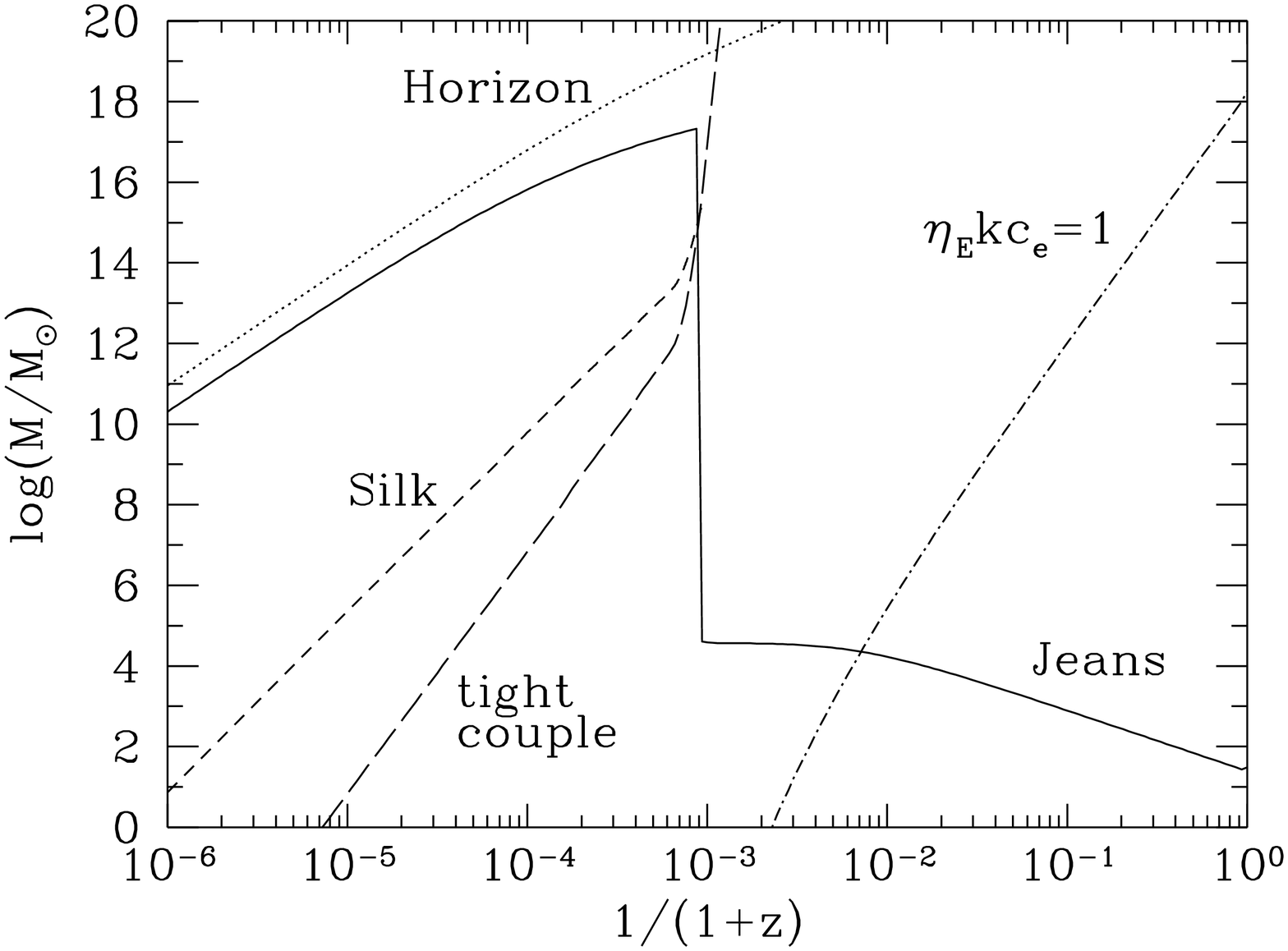,width=13cm}
  \end{center}
\caption{The physical mass scales for the baryon perturbations.
The cosmological parameters taken here are same as those in Fig.1
The definitions of these scales are summarized in appendix B.
In the figure, 
`Horizon' means the horizon scale, 
`Silk' does the diffusion damping scale,
`tight couple' does the breaking scale of the tight coupling
approximation of baryon and photon fluids,
`Jeans' does the Jeans scale.
}
\label{Fig.2}
\end{figure}

To get a physical insight, we refer to the familiar illustrative 
figure, Fig.2, which gives temporal variations of various physical 
sizes in the expanding universe. The definitions of the curves 
are summarized in appendix B. (See also ref.\cite{YSS}.)
The adiabatic index $\gamma$ after the recombination is changed 
when the Compton energy transfer time is equal to the
sound oscillation time scale.
In this figure, we show the line on which the two time scales
are equal, i.e., $\eta_{\rm E} k \ce=1$.
Note that we use the (baryon) mass scale in the unit of the solar mass 
instead of $k$, with employing the relation 
$M=(4\pi\rho_{\rm b}/3)( \pi a/k a_0)^3$.
The mass and the comoving wave number which satisfy the relation
$\eta_{\rm E} k \ce=1$ are expressed 
(see also appendix B), 
\begin{eqnarray}
  &&k={9.2 x^{(0)}(1-y_{\rm p}/2) 
  \over (\mu^3 T_{\rm b}^{(0)})^{1/2} }
  (1+z)^3
  ~\Mpc^{-1},
\\
  &&M={4.6\times 10^{10} (\mu^3 T_{\rm b}^{(0)})^{3/2}
               \over x^{(0)}{}^3 (1-y_{\rm p}/2)^3}
  (1+z)^{-9}
  \Omegab h^2 
  ~\Msolar ,
\end{eqnarray}
where we set the temperature of the microwave background
at present $T^{(0)}_\gamma(t_0)=2.726$K, $T_{\rm b}$ is the baryon 
temperature in unit of Kelvin, and $\mu=1-3y_{\rm p}/4$.

As is shown  in Fig.2, the Jeans scale after the recombination
has the plateau. In this stage the energy transfer
between background photons and the baryon fluid is effective
through the residual electrons, and the matter temperature
follows the photon temperature. As the universe expands, 
however, the energy transfer time rises above the 
Hubble expansion time. After that epoch the matter temperature
cools adiabatically and drops as $\Te\propto 1/a^{2}$.
The broken corner of the plateau is the critical time
that the two time scales become equal, where this epoch
is roughly estimated as 
$z\simeq 1000 (\Omega_b h^2)^{2/5}$ \cite{Peeb2} (see also appendix C).

\begin{figure}[t]
  \begin{center}
  \epsfile{file=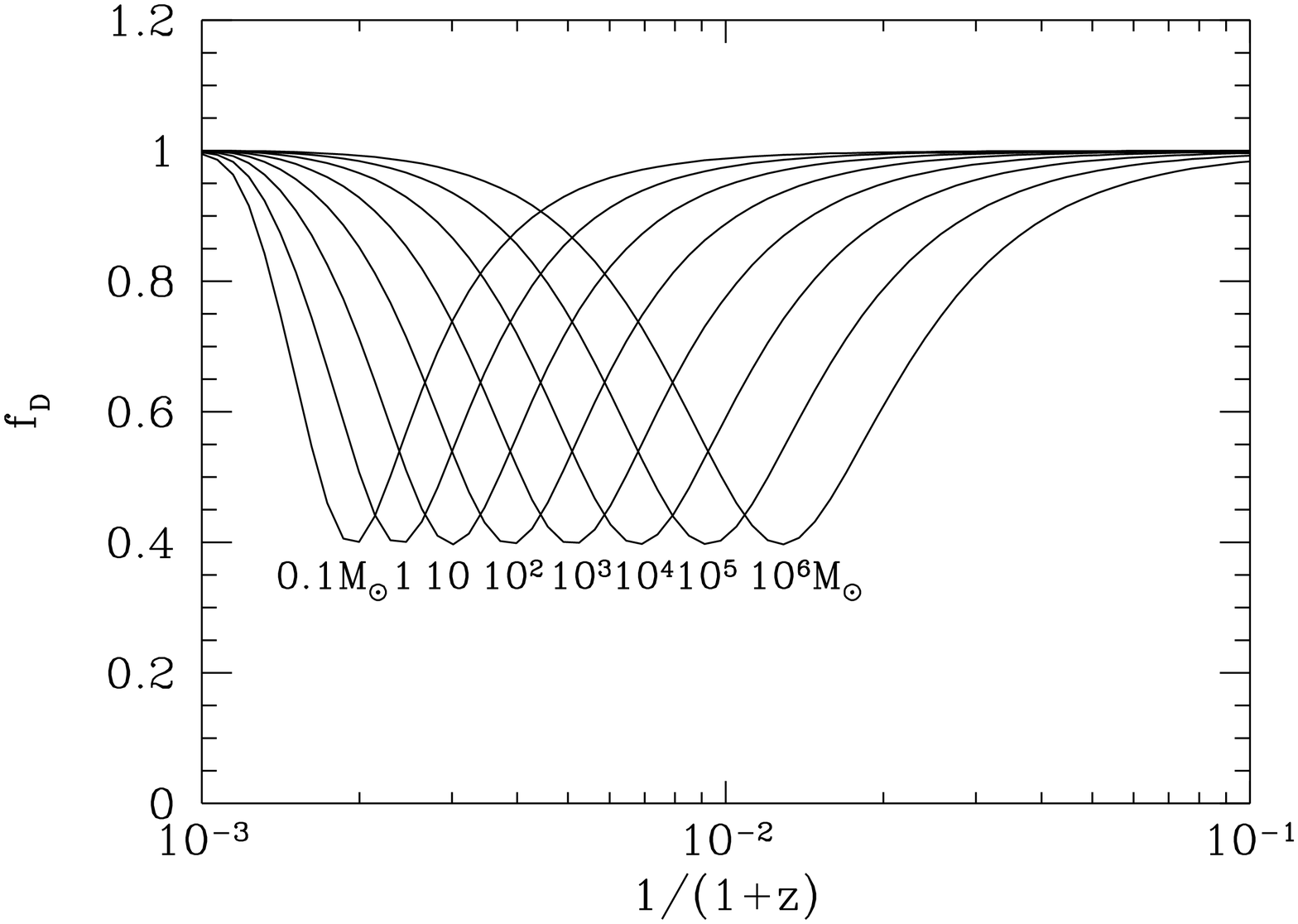,width=13cm}
  \end{center}
\caption{The behaviors of $f_D$ as the function of $1/(1+z)$. 
  Each lines are for the mass scales, $M=1\Ms$, $10^3\Ms$, $10^6\Ms$, 
  $10^9\Ms$, $10^{12}\Ms$, respectively.
  The cosmological parameters are same as those in Fig.1
  The damping phenomenon occurs when the $\gamma$
  is changed from $1$ to $5/3$.
}
\label{Fig.3}
\end{figure}

The line of $\eta_{\rm E} k \ce=1$ crosses at the broken
corner of the plateau of the Jeans mass scale.
This necessarily happens because of the following reason.
The Jeans scale is 
the scale at which the sound oscillation time is equal to 
the free fall time of the perturbation or the Hubble expansion 
time. Since the broken corner of the
plateau is the epoch when the Hubble expansion time 
is equal to the energy transfer time through the Compton interaction
between the background photons and the baryon-electron fluid,
the cross point of two lines is the point when the sound 
oscillation time, the Hubble expansion
time and the energy transfer time become all same.

Now let us discuss an interesting result derived from the 
dispersion relation (\ref{dispersionA}).
As it is apparent from the approximated solution, eq.(\ref{solapprox}),
$\omega$ generally has an imaginary part.
This imaginary part of $\omega$ represents the exponential 
damping of the wave oscillation if ${\rm Im}~\omega < 0$.
Since the period of the sound oscillation is 
$T=2\pi/\vert{\rm~Re}\omega\vert$, the damping factor for 
the amplitude during one period is
\begin{equation}
  f_D=\exp\biggl[-2\pi\bigg\vert{{\rm Im}~\omega\over{\rm Re}~\omega}
 \bigg\vert\biggr].
\end{equation}
We show the behavior of $f_D$ in Fig.3, 
in the similar way to Fig.1 as the function of $1/(1+z)$.
As is shown in Fig.3, this damping phenomenon may 
have an effect on the evolution of baryon perturbations.
We should be notice that our discussions here are based on the assumption 
that the solutions have the wave form.  
In other word, this damping process is effective only inside the Jeans
scale.

\begin{figure}[t]
  \begin{center}
  \epsfile{file=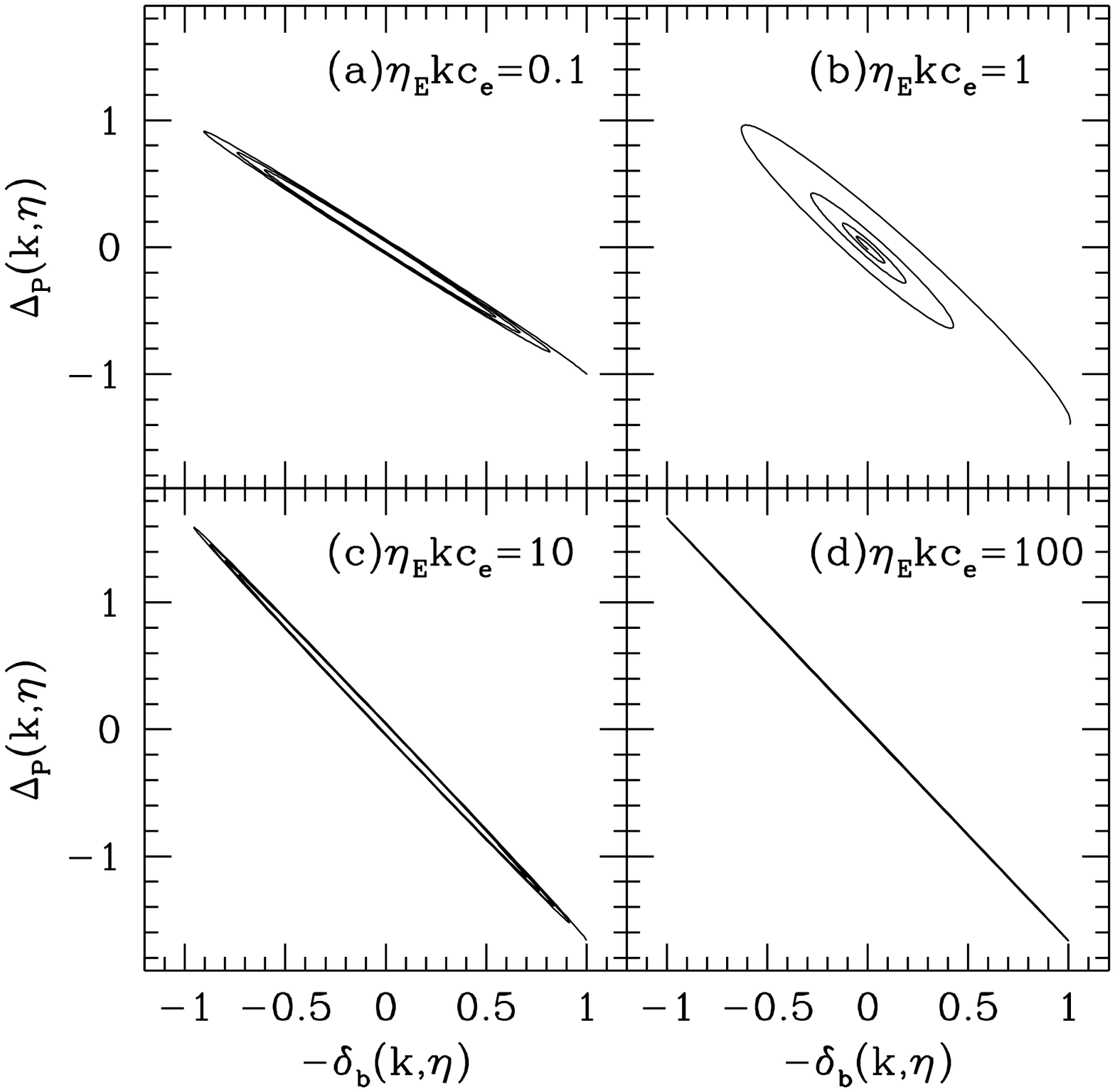,width=13cm}
  \end{center}
\caption{The trajectory of the solution on
$(\Delta_{P}-\delta_{{\rm b}})$-plane. 
The horizontal axis is the baryon density perturbation
$-\delta_{{\rm b}}(k,\eta)$,
and the vertical axis is the pressure perturbation 
$\Delta_{{\rm P}}(k,\eta)$.
We take $\eta_{\rm E} k \ce=0.1, 1, 10$ and $100$.
}
\label{Fig.4}
\end{figure}

In order to understand this damping mechanism, we 
show another aspect of the solutions of Eq.(\ref{cubiceqA}).
Introducing 
$\omega_{\rm R}={\rm Re} \omega$,
$\omega_{\rm I}={\rm Im} \omega$,
we take the solution
\begin{equation}
  \deltab(k,\eta)=-\cos(\omega_{\rm R}\eta)
  ~e^{-\omega_{\rm I}\eta}.
\label{solution}
\end{equation}
The initial value is $\deltab(k,\eta=0)=-1$.
Here $\omega$ is obtained by solving the 
dispersion relation Eq.(\ref{dispersionA}).
For $\eta_{\rm E} k \ce \gg 1 $ and $\eta_{\rm E} k \ce \ll 1$,
solutions are mere harmonic oscillations with no damping.
Therefore the solutions around $\eta_{\rm E} k \ce \simeq 1$
should be investigated.  
By employing Eqs. (\ref{deltabApplB}) and (\ref{eqst1}),
the trajectories of the solutions on 
$(\Delta_{P}-\delta_{{\rm b}})$-plane are shown in Fig.4.
The horizontal axis is $-\delta_{{\rm b}}(k,\eta)$
and the vertical axis is $\Delta_{P}(k,\eta)$.
The non-dimensional quantity $\eta_{\rm E} k \ce$ 
is the unique physical parameter of the equation.
We have chosen $\eta_{\rm E} k \ce=0.1,~1, ~10, {\rm and} ~100$ in
Fig.4.

From the relation $-\deltab=\delta V/V$,
where $V$ is the volume for unit particle number, 
the horizontal axis can be regarded as the change of volume 
per unit particle number.  Thus we can regard the trajectories
in Fig.4 in the similar way to the thermodynamical cycles
in (pressure-volume)-plane. 
The damping of the oscillation is in proportion to 
the deviation in one cycle.
The most significant damping occurs when $\eta_{\rm E} k \ce= 1$ (Fig.4(b)).
The damping does not happen if the deviation in one
cycle is negligible as we can see in Fig.4(d).

\section{Summary and Discussions}
We have formulated equations for a fluid system with
the electron, neutral and ionized hydrogen atoms and neutral helium 
atoms with taking into account 
the energy transfer between the background photons and 
the  residual electrons through the Compton interaction.
Using this formulation, we have studied the time evolution of 
the sound speed and the Jeans scale after the recombination.   
We found that 
the behavior of the adiabatic index $\gamma$ after the recombination 
is controlled by the ratio of the Compton energy transfer time scale to
the sound oscillation time scale. 
Then $\gamma$ (or the sound speed) depends on scales of the 
perturbations, and is changed from $1$ to $5/3$ 
in earlier stage for smaller scale perturbations.
This formalism enables us to calculate the linear evolution 
of the very small scale baryon density perturbations \cite{YSS}.

We have also discussed the small damping feature of
the sound oscillation when $\gamma$ changes from $1$
to $5/3$. 
This effect works inside the Jeans scale and seems to be 
negligible on scales  $M  \gsim 10^6 M_\odot$.
If early reionization occurs, however, the ionization fraction and 
the Jeans scale increase. And $\gamma$ changes from $5/3$ to $1$.  
The damping of the baryon perturbation would be effective on larger
scales during the process of the reionization of the universe.
It has been pointed out that the neutral gas cloud could have
an instability in a specific photon background \cite{Hogan}. 
Since we have not consider the specific situation in our
paper, an unstable mode does not appear in our equations. 
In the present paper we have also neglected perturbations to the 
ionization fraction. Future work may be required on this point.

It will be important to quantify the limitation of our formalism.
Our basic assumption is the fact that the baryon-electron system 
is treated as tightly coupled single-fluid. 
This assumption becomes not being correct when the collision 
time scale of neutral interaction rises above the Hubble 
expansion time. 
This epoch can be estimated in the following way \cite{Blanchard}.
The matter temperature can be written as
$ T_{\rm b}=4.5\times 10^{-3}(\Omegab h^2)^{-2/5} (1+z)^2$[K], 
after the energy transfer through the Compton interaction 
becomes ineffective (see appendix C).
Then the mean free time for collision between neutral hydrogen
atoms is $t_c\simeq 1/(n_{\rm B}\sigma v)$, where $\sigma$ is
the cross section of neutral atoms $\sigma\simeq \pi r_B^2$
($r_B$ is the Bohr radius), and $v=\sqrt{3T_{\rm b}/m_{\rm H}}$.
The ratio of $t_c$ to the Hubble expansion time is therefore expressed as
$t_c H\simeq3.1(\Omega_0h^2)^{1/2}(\Omega_bh^2)^{-4/5}(1+z)^{-5/2}$.
Eventually we get the red-shift at $t_c H=1$ as 
$(1+z)\simeq1.6~(\Omega_0h^2)^{0.2}
(\Omegab h^2)^{-0.32}$. This is small enough as long as we
consider the linear stage of the density perturbations.

Numerical simulations and semi-analytic calculations 
of structure formation employ the linear matter
power spectrum as their initial conditions.
This linear matter spectrum is usually calculated 
without taking into account 
the baryon pressure term after the recombination.
It is appropriate, however, 
only if the structures which are larger than the Jeans
scale are considered.  According to the hierarchical clustering scenario,
smaller objects are formed earlier than larger ones.  
We expect the scale of the first collapsing object is very close to the
Jeans scale.  Therefore very
accurate estimate of the matter spectrum including the baryon pressure 
term is necessary to understand the early formation of bound 
objects \cite{YSS}.
Once the first collapsing 
objects are formed, 
they may cool down through formation of hydrogen molecules and may
fragment into smaller radiating objects like stars and/or quasars 
\cite{smt,mts,susa,RO}.
The first radiating objects formed Str\"omgren sphere around 
them \cite{sato1}
and may eventually 
ionize all the surrounding gas by UV radiation.  This reionization
process changes the Jeans scale \cite{CSK}.  
After that, full numerical
or semi-analytic calculations including collapsing objects are required.

\vspace{0.3mm}
\begin{center}
{\bf ACKNOWLEDGMENT}
\end{center}
K.Y. thanks K.Kusano, Y.Kojima and N.Ohuo for comments at the early 
stage of this work. We thank R.Nishi for comments. 
This work is partially supported by the 
Grants-in-Aid for Scientific Research of Ministry of Education, Science
and Culture of Japan (Nos.~09740203 and 09440106).

\vspace{0.3mm}


\begin{appendix}

\section{Rate equation}
In order to complete the fluid equations we should add the rate equation.
For simplicity, we neglect the helium fraction, i.e., 
$\nb=n_{\rm H}$, and neglect the helium recombination process 
in the rate equation. Then, the rate equation is \cite{Peeb2}
\begin{equation}
  {\partial x\over\partial\eta}=-\bigl<\sigma v\bigr> {\aova}
  \nb C\biggl(x^2-(1-x)
  {\alphaeq^2\over1-\alphaeq}\biggr),
\label{rateeqA}
\end{equation}
where $\bigl<\sigma v\bigr>$ is the rate coefficient for recombination 
to excited states, 
$\alphaeq$ is the equiliblium ionization fraction 
which is given by the Saha equation 
\begin{equation}
  {\alphaeq^2\over1-\alphaeq}={\mp\over\rhob}
  \biggl({\me \Te\over2\pi}\biggr)^{3/2}
  e^{-13.6{\rm eV}/\Te},
\label{SahaeqA}
\end{equation}
and  
\begin{eqnarray}
  C={1+K\Lambda n_{\rm 1s}\over1+K(\Lambda+\beta_{\rm e})n_{\rm 1s}} ~.
\end{eqnarray}
Here $n_{\rm 1s}$ is the number density of 
hydrogen in the electron ground state, for which we approximate 
$n_{\rm 1s}=\nb(1-x)$, 
$\Lambda$ is the decay rate from the excited state,
$\beta_e=\bigl<\sigma v\bigr>(\me\Te/2\pi)^{3/2}e^{-3.4~{\rm eV}/Te}$ and
$K=(a/{\dot a})\lambda_\alpha^3/8\pi$ 
with $\lambda_\alpha$ being the Lyman alpha photon wave length.

The Zero-th order equations of the rate equation and the Saha
equation are the same forms of Eqs.(\ref{rateeqA}) and 
(\ref{SahaeqA}) replaced the variables with the
zero-th order quantities.

Next we consider the perturbation of the rate equation.
We define
\begin{eqnarray}
  &&\alphaeq=\alphaeq^{(0)}(\eta)+\delta\alphaeq(\eta,\bfx),
\end{eqnarray}
then
\begin{eqnarray}
  &&{d\delta x\over d\eta}={a\over a_0}\bigl<\sigma v\bigr>\nb^{(0)} C
  \biggl[\biggl({x^{(0)}}^2-{(1-x^{(0)}){\alphaeq^{(0)}}^2\over1-\alphaeq^{(0)}}\biggr)
  \biggl({\delta C\over C}+\deltab\biggr)
\nonumber
\\
  &&\hspace{4cm}
  +\biggl(2x^{(0)}+{{\alphaeq^{(0)}}^2\over1-\alphaeq^{(0)}}\biggr)\delta x
  -{(1-x^{(0)})
  \alphaeq^{(0)}(2-\alphaeq^{(0)})\over(1-\alphaeq^{(0)})^2}\delta\alphaeq
  \biggr],
\end{eqnarray}
where
\begin{equation}
  {\delta C\over C}
  ={-K\beta_e\nb^{(0)}\Bigl((1-x^{(0)})\deltab-\delta x\Bigr)\over
  \Bigl(1+K\Lambda\nb^{(0)}(1-x^{(0)})\Bigr)
  \Bigl(1+K(\Lambda+\beta_e)\nb^{(0)}(1-x^{(0)})\Bigr)} ~ .
\end{equation}
We have assumed that $\bigl<\sigma v\bigr>$ and
$\beta_e$ are constant.

The perturbation of the Saha equation can be written as
\begin{equation}
  \delta\alphaeq=
  {\partial\alphaeq(P^{(0)},\rhob^{(0)})\over
  \partial\rhob^{(0)}}\bigg|_{P^{(0)}}\rhob^{(0)}\deltab
  +{\partial\alphaeq(P^{(0)},\rhob^{(0)})\over
  \partial P^{(0)}}\bigg|_{\rhob^{(0)}}P^{(0)}\Delta_P.
\end{equation}

\section{Physical scales}

In this appendix, we summarize the physical scales which
are important for the evolution of the baryon perturbations
on small scales. 
The definitions of the physical scales in Fig.2 are given.
Mass scales are defined by the amount of baryonic components inside
the systems.
Here we have set that the temperature of the microwave background
at present $T_\gamma(t_0)=2.726K$.
We write $f_\nu$ as the neutrino fraction of the energy 
density in the massless particles, and $f_\nu=0.405$ 
in case of the standard three families of massless neutrinos.

First of all, we use the notation for the physical wave number 
(length) and the comoving wave number (length) as 
\begin{equation} 
  \kcomv=\Bigl({a\over a_0}\Bigr)\kphys,
  \hspace{5mm}
  \lambda^{\rm comv}=\Bigl({a_0\over a}\Bigr)  \lambda^{\rm phys} 
  = 2\pi /\kcomv.
\end{equation}
It will be useful to give the relation between the red-shift 
$z$ and the scale factor normalized at the matter radiation 
equality $a_{\rm eq}$,
\begin{equation}
  {a\over a_{\rm eq}}=4.04\times10^{4}(1-f_\nu)\Omega_0 h^2(1+z)^{-1}.
\end{equation}

\subsection{ Horizon Scale } 
We define the horizon wave number and the horizon mass as
\begin{eqnarray}
  && {1\over k_{\rm H}^{\rm comv}}=\eta,
\\
  &&M_{\rm H}
  ={4\pi\rhob\over 3}\biggl({\lambda_{\rm H}^{\rm phys}\over2}\biggr)^3
  ={4\pi\rhob\over 3}\biggl({\pi\over k_{\rm H}^{\rm phys}}\biggr)^3,
\end{eqnarray}
which derive
\begin{eqnarray}
  && k^{\rm comv}_{\rm H}=3.35\times10^{-2} 
  \biggl(\sqrt{1+{a\over \aeq}}-1\biggr)^{-1}
  (1-\fnu)^{1/2} 
  \Omegam h^2
  ~\Mpc^{-1},
\\
  &&M_{\rm H}
  =9.57\times10^{17} 
  \biggl(\sqrt{1+{a\over \aeq}}-1\biggr)^{3}
  (1-\fnu)^{-3/2} 
  \Omegab h^2 (\Omegam h^2)^{-3} \Msolar.
\end{eqnarray}

\subsection{Jeans scale before the recombination }
We define the Jeans wave length (wave number) and the Jeans mass as
\begin{eqnarray}
  &&\lambda_{\rm J}^{\rm phys}={2\pi\over k_{\rm J}^{\rm phys}}=
  \sqrt{{\pi \cs^2\over G(\rho_{\rm m}+\rho_\gamma)}},
\\
  &&\MJ
  ={4\pi\rhob\over 3}\biggl({\lambda_{\rm J}^{\rm phys}\over2}\biggr)^3
  ={4\pi\rhob\over 3}\biggl({\pi\over k_{\rm J}^{\rm phys}}\biggr)^3,
\end{eqnarray}
where
\begin{equation}
  \cs^2={1\over 3 (1+R)},
\end{equation}
and 
$\rho_{\rm m}=\rhob+\rho_{\rm dm}$ with $\rho_{\rm dm}$ being the
energy density of the dark component.
Then we have
\begin{eqnarray}
  &&k_{\rm J}^{\rm comv}=1.42\times10^{-1} 
  \biggl({\aeq\over a}+(1-\fnu){\aeq^2\over a^2}\biggr)^{1/2}
  (1+R)^{1/2}
  (1-f_\nu)^{1/2}
  \Omegam h^2
  ~\Mpc^{-1},
\\
  &&{\MJ}=
\nonumber
\\
  &&\hspace{0.5cm}
  1.25\times10^{16}
  \biggl({\aeq\over a}+(1-\fnu){\aeq^2\over a^2}\biggr)^{-3/2}
  (1+R)^{-3/2}
  (1-f_\nu)^{-3/2}
  (\Omegam h^2)^{-3}\Omegab h^2
  ~\Msolar.
\end{eqnarray}

\subsection{Jeans scale after the recombination}
We can define the Jeans scale after the recombination as 
\begin{equation}
  \lambda_{\rm J}^{\rm phys}={2\pi\over k_{\rm J}^{\rm phys}}=
  \sqrt{{\pi \cs^2\over G\rho_{\rm m}}},
\end{equation}
with
\begin{equation}
  \cs^2=\gamma {P^{(0)}\over \rhob^{(0)}}
  = 9.18\times 10^{-14}\gamma \mu   \Te^{(0)},
\end{equation}
where $\mu=1-3y_{\rm p}/4$  for  $x^{(0)}\ll 1$,
and $\Te$ is the matter temperature in unit of Kelvin. 
Then we have 
\begin{eqnarray}
  &&k_{\rm J}^{\rm comv}=2.71\times10^{5}
  \biggl(\gamma\mu\Te {a \over \aeq}\biggr)^{-1/2}
  (1-f_\nu)^{1/2}
  \Omegam h^2
  ~\Mpc^{-1},
\\
  &&\MJ=1.81\times10^{-3} 
  \biggl(\gamma\mu\Te {a\over \aeq}\biggr)^{3/2}
  (1-f_\nu)^{-3/2}
  (\Omegam h^2)^{-3}\Omegab h^2
  ~\Msolar.
\end{eqnarray}

\subsection{Diffusion damping scale} 
We define the diffusion damping scales as \cite{HSI}
\begin{eqnarray}
  &&\left({1\over k_D^{\rm comv}}\right)^2={1\over6}\int {1\over\dot\tau}
  d\eta{R^2+4(1+R)/5\over(1+R)^2},
\\
  &&{M_{\rm D}}=
  {4\pi\over 3}\rhob\biggl({\pi\over k_{\rm D}^{\rm phys}}\biggr)^3,
\end{eqnarray}
where $\dot\tau=n_e\sigma_T(a/a_0)$. For $R\ll1$, we get  
\begin{eqnarray}
  &&k_{\rm D}^{\rm comv}
  =1.38\times 10^{2} 
  \biggl({u(a/\aeq)\over x^{(0)}(1-y_p/2)}\biggr)^{-1/2}
  (1-f_\nu)^{5/4} 
  (\Omegam h^2)^{3/2}(\Omegab h^2)^{1/2}
  ~\Mpc^{-1},
\\
  &&M_{\rm D}=
  1.38\times10^{7}
  \biggl({u(a/\aeq)\over x^{(0)}(1-y_p/2)}\biggr)^{3/2}
  (1-f_\nu)^{-15/4} 
  (\Omegam h^2)^{-9/2}(\Omegab h^2)^{-1/2}
  ~{\Msolar},
\end{eqnarray}
where $u(y)=(\sqrt{1+y}(16-8y+6y^2)-16)/15$.

\subsection{Breaking Scale of the tight coupling approximation}
The breaking scale of the tight coupling approximation is
defined by $1/k^{\rm phys}_{\rm BR}=1/n_e\sigma_T$, i.e.,
\begin{eqnarray}
  && k_{\rm BR}^{\rm comv}=n_e\sigma_T {a\over a_0},
\\
  && M_{\rm BR}=
  {4\pi\over 3}\rhob\biggl({\pi\over k_{\rm BR}^{\rm phys}}\biggr)^3.
\end{eqnarray}
Then we have
\begin{eqnarray}
  &&k^{\rm comv}_{\rm BR}=3.77\times 10^{4}
  \biggl({(a/\aeq)^2\over x^{(0)}(1-y_p/2)}\biggr)^{-1}
  (1-f_\nu)^{2} 
  (\Omegam h^2)^{2}(\Omegab h^2)
  ~\Mpc^{-1},
\\
  && {M_{\rm BR}}
  =6.75\times10^{-1}
  \biggl({(a/\aeq)^2\over x^{(0)}(1-y_p/2)}\biggr)^{3}
  (1-f_\nu)^{-6} 
  (\Omegam h^2)^{-6}(\Omegab h^2)^{-2}
  ~{\Msolar}.
\end{eqnarray}

\subsection{$\gamma$ transition epoch after the recombination}

The adiabatic index $\gamma$ after the recombination is changed 
when the Compton energy transfer time is equal to the
sound oscillation time scale. We define this scale 
by $1/k_{\rm c}^{\rm comv}=\ce \eta_{\rm E}$, which leads
\begin{eqnarray}
  &&k^{\rm comv}_{\rm c}={6.08\times 10^{14} x^{(0)} 
  (1-y_{\rm p}/2)  (1-f_\nu)^{3}
  \over (\mu^3 T_{\rm e})^{1/2} (a/\aeq)^3}
  (\Omega_0 h^2)^3
  ~\Mpc^{-1},
\\
  &&M_{\rm c}={1.60\times 10^{-31} (\mu^3 T_{\rm e})^{3/2} (a/\aeq)^9
               \over x^{(0)}{}^3 (1-y_{\rm p}/2)^3 (1-f_\nu)^9}
  \Omegab h^2 (\Omega_0 h^2)^{-9}
  ~\Msolar,
\end{eqnarray}
where $\Te$ is the matter temperature in unit of Kelvin as 
is mentioned above.

\section{ Matter Temperature }
As we have derived in \S V, the matter temperature follows 
\begin{equation}
 \Te^{(0)}{}'+2{a'\over a}\Te^{(0)}
  =\eta_{\rm E}^{-1}(T^{(0)}_\gamma-\Te^{(0)}),
\end{equation}
where $T_\gamma^{(0)}$ is the photon temperature,
and the Compton energy transfer time scale $\eta_{\rm E}$ 
is defined as 
\begin{equation}
  \eta_{\rm E}^{-1}={8\over3}{\aova}{x^{(0)}\Bigl(1-\yp/2\Bigr)
  \sigma_{\rm T}\rho_\gamma^{(0)}\over\me
  \Bigl(1+x^{(0)}-(x^{(0)}+3/2)\yp/2\Bigr)} ~.
\label{etaEinv}
\end{equation}
The formal solution is 
\begin{equation}
  \Te^{(0)}={1\over a^2}\int_0^\eta d \eta' ~a^2
  T^{(0)}_\gamma\eta_{\rm E}^{-1}
  \exp\biggl[-\int_{\eta'}^\eta d\eta'' \eta_{\rm E}^{-1} \biggr].
\end{equation}
We can obtain the epoch when the matter temperature deviates from
the photon temperature as follows. 
This epoch is naturally defined as the epoch 
when the function $\eta_{\rm E}^{-1} 
\exp\bigl(-\int_{\eta}^{\eta_0} d\eta' \eta_{\rm E}^{-1}\bigr)$ 
takes its maximum value.
Thus we need to solve the equation
$\dot\eta_{\rm E}^{-1}+(\eta_{\rm E}^{-1})^2=0$.
From Eq. (\ref{etaEinv}), 
\begin{equation}
  \eta_{\rm E}^{-1}=1.8\times10^8(1-f_\nu)^3( \Omega_0 h^2)^3 
  {x^{(0)}\bigl(1-\yp/2\bigr)\over \bigl(1-3\yp/4\bigr) } 
  \biggl({a\over a_{\rm eq}}\biggr)^{-3} ~{\rm Mpc}^{-1} .
\end{equation}
Therefore we get the epoch as 
\begin{equation}
  {a\over\aeq}=2.3\times10^3 x^{(0)}{}^{2/5}(\Omega_0h^2)^{4/5},
\label{tempdecA}
\end{equation}
with neglecting the time variation of the ionization fraction $x^{(0)}$.
Here we  set $y_{\rm p}=0.23$ and  $f_{\nu}=0.405$.

We approximate the fraction of the residual electron as \cite{Bernstein}
\begin{equation}
  x^{(0)}\simeq 10^{-5}(\Omega_0h^2)^{1/2} (\Omegab h^2)^{-1} .
\end{equation}
Then from Eq.(\ref{tempdecA}), we conclude that the matter temperature 
decouples from the photon's at $(1+z)\simeq1000 (\Omegab h^2)^{2/5}$.
After this epoch, the matter temperature is adiabatically cooling.

According to the fully numerical calculation \cite{HSI}, 
the matter temperature in this adiabatic cooling phase is well 
reproduced by the formula,
\begin{equation}
  T_{\rm b}^{(0)}= 4.5\times 10^{-3} (1+z)^2 (\Omegab h^2)^{-2/5} ~{\rm K}.
\label{Tbanaly}
\end{equation}

\end{appendix}


\begin{references}
\bibitem{silk}
J. Silk, Astrophys. J. {\bf 151}, 459 (1968).
\bibitem{py}
P.J.E Peebles \& J.T. Yu, Astrophys. J. {\bf 162}, 815 (1970).
\bibitem{smt}
H. Sato, T. Matsuda \& H. Takeda, Prog. Theor. Phys. Suppl. {\bf 49}, 
11 (1971).
\bibitem{wein}
S. Weinberg, {\it Gravitation and Cosmology},
(John Wiley \& Sons, Inc., New York, 1972).
\bibitem{Peeb1}
Peebles, P. J. E., {\it The Large Scale Structure in the Universe}, 
(Princeton University Press, Princeton, 1980).
\bibitem{sphr}
W.H. Press \& P. Schechter, Astrophys. J. {\bf 187}, 425 (1974);
J.R. Bond, S. Cole, G. Efstathiou \& N. Kaiser, Astrophys. J. {\bf 379},
440 (1991).
\bibitem{HSS}
W. Hu, D. Scott, \& J. Silk,  Phys. Rev. D {\bf 49}, 648 (1994).
\bibitem{DJ}
S. Dodelson \& M. Jubas, Astrophys. J. {\bf 439}, 503 (1995).
\bibitem{MS}
F. Mandel, \& G. Shaw, {\it Quantum Field Theory } (John Wiley \&
Sons Inc., New York, 1984).
\bibitem{Peeb2}
P.J.E. Peebles, {\it Principles of Physical Cosmology},
(Princeton University Press, Princeton, 1993).
\bibitem{JW}
B.J.T. Jones \& R.F.G. Wyse, Astronomy and Astrophys.{\bf 149 }, 144 (1985).
\bibitem{HSI}
W. Hu \& N. Sugiyama, Astrophys. J. {\bf 444 }, 489 (1995).
\bibitem{YSS}
K. Yamamoto, N. Sugiyama, \& H. Sato, preprint, in preparation (1997).

\bibitem{Hogan}
C. J. Hogan, Nature {\bf 350}, 469 (1991) ; {\bf 359}, 40 (1992).
\bibitem{Blanchard}
A. Blanchard, in {\it Galaxy Formation}, ed. by J.Silk and N. Vittorio,
North-Holland, (1994).
\bibitem{mts}
T. Matsuda, H. Takeda \& H. Sato, Prog. Theor. Phys. {\bf 42}, 
219 (1969).
\bibitem{susa}
H. Susa, H.Uehara \& R. Nishi, Prog. Theor. Phys. {\bf 76}, 
1073 (1996);
H. Uehara, H. Susa, R. Nishi,  M. Yamada, \& T. Nakamura, 
Astrophys. J. Lett. {\bf 473}, L95 (1996).

\bibitem{RO}
M.J. Rees \& J.P. Ostriker, Mon. Not. R. astr. Soc. {\bf 179 }, 541 (1977).
\bibitem{sato1}
H. Sato, Prog. Theor. Phys. {\bf 92}, 
37 (1994); Z. Haiman, A. Loeb, preprint (astro-ph/9611028).
\bibitem{CSK}
T. Chiba, N. Sugiyama, \& M. Kawasaki, preprint.
\bibitem{Bernstein}
J. Bernstein, {\it Kinetic Theory in the Expanding Universe},
Cambridge University Press, (1988).
\end{references}
\end{document}